\documentclass{elsart}
\usepackage{graphicx}
\usepackage{amssymb}
\usepackage{amsfonts}
\begin{document}
\begin{frontmatter}

\title{
ANTIFERROMAGNETIC EFFECTS IN CHAOTIC MAP LATTICES WITH A
CONSERVATION LAW}
\author{Leonardo Angelini}
\address{Dipartimento Interateneo di Fisica and Istituto Nazionale di Fisica Nucleare}
\address{via Amendola 173, 70126 Bari, Italy}
 \maketitle
\begin{abstract}
Some results about phase separation in coupled map lattices
satisfying a conservation law are presented. It is shown that this
constraint is the origin of interesting antiferromagnetic
effective couplings and allows transitions to antiferromagnetic
and superantiferromagnetic phases. Similarities and differences
between this models and statistical spin models are pointed out.
\end{abstract}
\end{frontmatter}

\section{Introduction}

Coupled map lattices (CMLs) are spatially extended dynamical systems that have been
considerably investigated recently on various grounds \cite{workshop}. They are approximations
to continuous systems particularly well suited for numerical and analytical calculations. From
another point of view they can be considered as phenomenological models describing the dynamics
of a large number of macroscopic chaotic subsystems. In both cases one is interested to study
their behaviour at length scales larger than the scale where chaos is present.

Particular attention has been dedicated to the study of collective behaviour in CMLs in
presence of conservation laws. This class of models are considered relevant to describe several
physical phenomena like surface waves in a container or disturbances in fluids, where mass,
momentum and energy are conserved. A motivation for their interest is that, as it can be easily
argued, the existence of conservative quantities should play an important role in determining
the long distance properties in such systems. Several authors
\cite{Grig-Cross,Bour-Cross,Grin-He-Ja-Bol} studied coupled map models with conservation laws
as possible models for chaos in extended systems, examining phase transitions from ordered to
chaotic states that are driven by the map parameters or by the value of the conserved quantity.

CMLs have been studied \cite{Mil-Hus,Mar-Cha-Man,Lem-Cha,Sch-Jus-Kan} also from another point
of view. In this case authors try to transfer concepts and results from equilibrium statistical
mechanics to systems, like CMLs, whose dynamics is microscopically irreversible and does not
satisfy detailed balance. In particular, studying systems of chaotic maps that exhibit a Ising
- like symmetry, it can be shown that they undergo, in particular conditions, a phase -
ordering dynamics. At large scale lengths the evolution of these deterministic systems has
coarsening properties similar to Ising models or their continuous versions, the time -
dependent Ginzburg - Landau equations \cite{Gunton,Iba-Gar-Tor-San}; the local chaos plays the
role of the stochastic noise present in a heat bath or in the noise term of the Langevin
equation. One expects obviously that, in the conserved order parameter case, this kind of CMLs
should show a behaviour similar to the so called model B. This model is described by the Cahn -
Hilliard \cite{Cah-Hil} equation, leading to asymptotic state in which the phases occupy two
large domains separated by a single boundary. Phase separation in CMLs with conserved dynamics
was observed \cite{APS99} in the case of a lattice of chaotic maps in contact with a thermal
bath evolving with Kawasaki dynamics.  In this case the temperature plays the role of an
addictive noise, while chaos produces chaotically fluctuating couplings. It was definitely
shown by J. Kockelkoren and H. Chat\'e \cite{Koc-Cha} that the coarsening process for this
model is strictly similar to that of the corresponding Ising model. In the same article the
authors proposed to study conserved dynamics of CMLs using the approach introduced by Y. Oono
and S. Puri \cite{Oon-Pur}. In this approach the dynamics corresponds to the discretization of
the Cahn - Hilliard equation, each map on a site of the lattice representing the effect of a
coarse - grained free energy. It was also shown that this model undergoes a phase transition
between a short range and a long range ordered phase.

The purpose of this paper is to extend the analysis of this model and to study the effect on
the phase separation process of the coordination number used to discretize Laplacian operators
on the lattice. In particular, it will be shown that the effect of the second Laplacian derived
from the Cahn - Hilliard equation, assuring the order parameter conservation, amounts to the
introduction of an effective antiferromagnetic coupling. The competition between ferromagnetic
and antiferromagnetic couplings in these systems plays an important role in the phase
separation process.

\section{The model}

Following \cite{Koc-Cha,Oon-Pur} we consider a two-dimensional square lattice of coupled
identical maps $f$ acting on real variables $x_i$. The discrete - time dynamics is governed by
the following equations:
\begin{eqnarray}
{\mathcal F}(x_i^t)&=&(1-{\mathcal N}g) f(x_i^t) + g {\sum_j}^i
f(x_j^t)
\label{dyn1}\\
 x_i^{t+1}&=&{\mathcal F}(x_i^t)-\frac{1}{{\mathcal
N}} {\sum_j}^i({\mathcal F}(x_j^t)-x_j^t), \label{dyn2}
\end{eqnarray}
where ${\mathcal N}$ is the chosen number of neighbors of site $i$, the sum is over these
neighbors and $g$ is the coupling strength; periodic boundary conditions are assumed. The
conservation of the order parameter
\begin{equation}
M=\sum_i x_i \label{ordpar}\end{equation} is incorporated by eq. (\ref{dyn2}),
representing the second Laplacian in the Cahn - Hilliard equation.

The map used in the numerical simulations is the following:
\begin{equation}
f(x)= \left \{
\begin{array}{cl}
 -{\mu\over 3}\exp{[ \alpha(x+{1\over 3})}] & if\;\; x\in[-\infty,-{1\over
 3}],
\\
 \mu x & if\;\; x\in [-{1 \over 3},{1 \over 3}], \\
 {\mu\over 3}\exp{[ \alpha({1\over 3}-x)}] & if\;\; x\in[{1\over
 3},+\infty],
\end{array}
\right. \label{map}
\end{equation}
i.e. a modified version of the map used in \cite{Koc-Cha}, that was defined on
the interval $[-1,1]$. The map \ref{map} is defined for every $x\in R$; the
modification is motivated by the fact (already stressed in
\cite{Grin-He-Ja-Bol} and verified by the author) that, due to the
redistribution step of the Oono - Puri dynamics (\ref{dyn2}), variables
$x_i(t)$ are not constrained to take value in $[-1,1]$. Details on this map can
be found in \cite{APS01}, where it was used for similar motivations. Choosing
$\mu =1.9$ and $\alpha =6$, $f$ has two symmetric chaotic attractors, one with
$x>0$ and the other with $x<0$; this allows the unambiguous definition of Ising
spin variables $\sigma_i^t = {\rm sgn}[x_i^t]$ associated to each dynamical
system.

To study the phase separation process, uncorrelated initial conditions were generated as
follows: one half of the sites were chosen at random and the corresponding values of $x$ were
assigned according to the invariant distribution of the chaotic attractor with $x>0$, while to
the other sites were similarly assigned values with $x<0$. With a good approximation the order
parameter $M$ vanishes. We associated an Ising spin configuration $\{s_i(t)\}= \{{\rm sgn}
[x_i(t)]\}$ to each configuration of the $x$ variable. Lattices from $256 \times 256$ up to
$512 \times 512$ with periodic boundary conditions were used. The average domain size $R(t)$
was measured by the relation $C[R(t),t]=1/2$, where $C(r,t)=\langle s_{i+r}(t) s_i(t)\rangle$
is the two point correlation function of the spin variables. $R(t)$ was averaged over many
different samples of initial conditions. More complicate correlation functions and related
lengths, that will be introduced in the following, have been measured. Another variable is
often considered in growth processes: the persistence $p(t)$ \cite{Der-Bra-God}, defined as the
fraction of sites that have not changed their initial $s$ values.  It has not been evaluated
because, to get a reliable calculation of persistence, one should use very large lattices which
is not compatible with a fine scan of the coupling variable.

One normally expects that, for large couplings, the conservation of the total "Magnetization"
(\ref{dyn2}) leads to equilibrium states where there are only two large domains with aligned
spins. However this conservation law is also compatible with more complicate phases. In
statistical spin models, they are generated by the presence of repulsive couplings in the
hamiltonian. In the sequel it will be shown that eq. (\ref{dyn2}), in addition to the
conservation law, generates these new couplings.

\section{Nearest neighbors simulations}
\begin{figure}
\begin{center}
\includegraphics*[width=12cm]{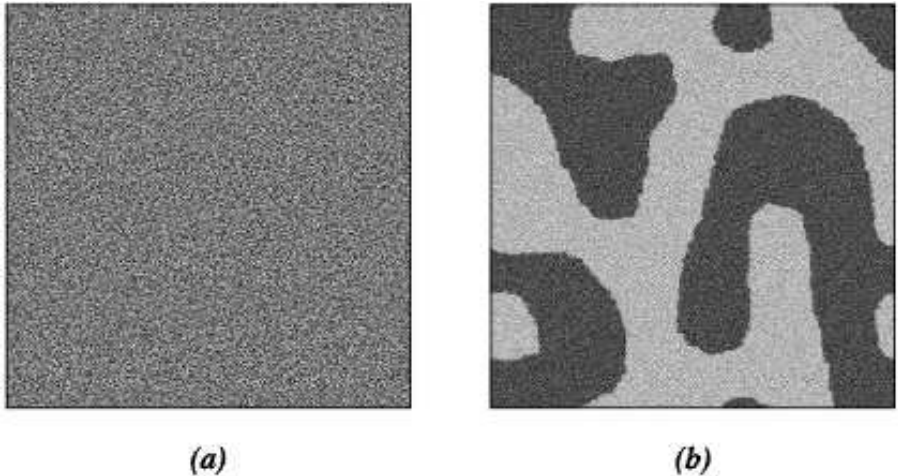}
\caption{Snapshots of a $256\times 256$ sites CML with four neighbors for
$g=0.01$ (a) and $g=0.05$ (b) after $10^6$ time steps. The full range of maps'
values has been colored with a 16 gray levels scale from black to white.}
\label{snap4f}
\end{center}
\end{figure}

\begin{figure}
\begin{center}
\includegraphics*[width=14cm]{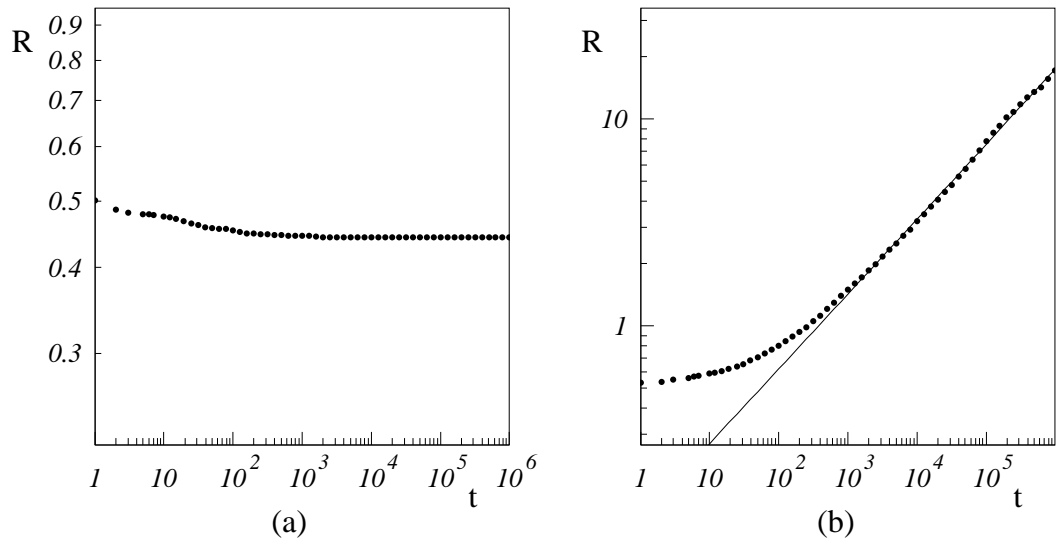}
\caption{Time evolution of the domain size $R(t)$ for $g=0.01$ (a) and $g=0.05$
(b). The solid lines is a best fit to the late time growth with power law
$R(t)= At^z$ with $z=0.341$.} \label{R4f}
\end{center}
\end{figure}

\begin{figure}
\begin{center}
\includegraphics*[width=14cm]{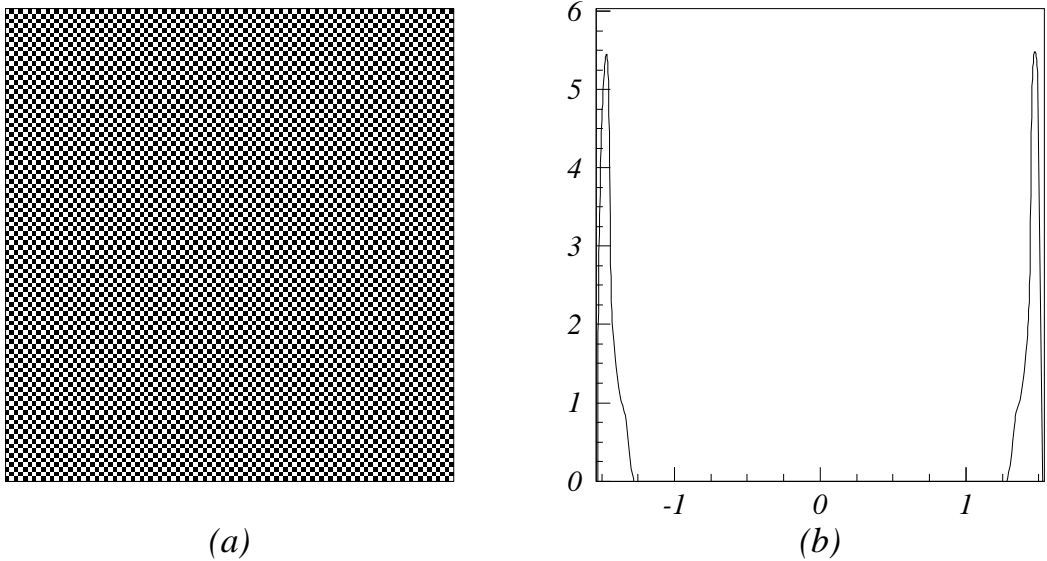}
\caption{The antiferromagnetic ordering at $g=0.128$ for a CML with four
neighbors: a) a magnified portion ($100 \times 100$) of a $512 \times 512$
lattice after $2000$ time steps, b) the probability distribution function for
the same lattice. The full range of maps' values has been colored with a 16
gray levels scale from black to white.} \label{n4g0128}
\end{center}
\end{figure}

\begin{figure}
\begin{center}
\includegraphics*[width=14cm]{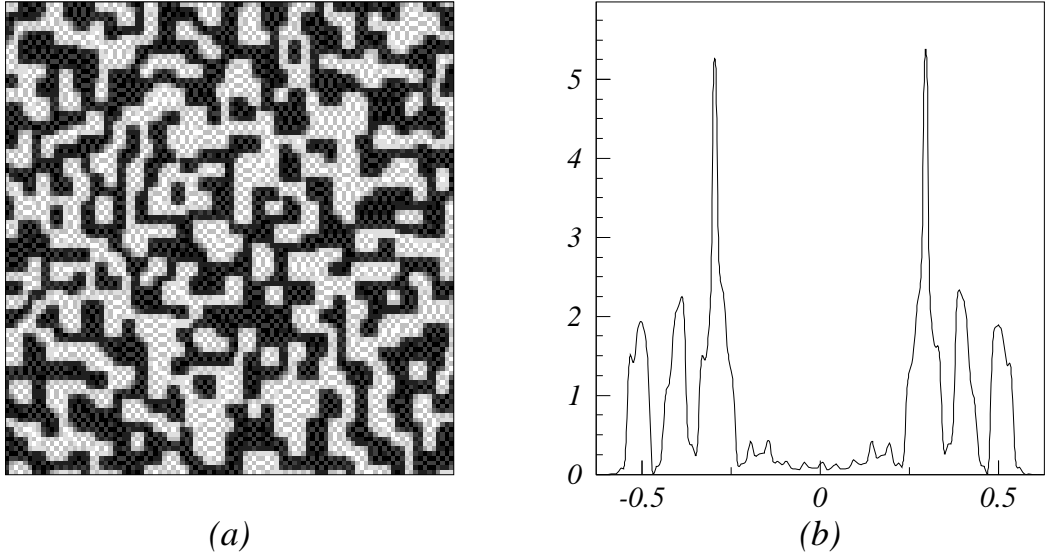}
\caption{Pre-transitional checkerboard structure at $g=0.127$: a) a magnified
portion ($100 \times 100$) of a $256 \times 256$ lattice after $10^6$ time
steps, b) the probability distribution function for the same lattice. The full
range of maps' values has been colored with a 16 gray levels scale from black
to white.} \label{n4g0127}
\end{center}
\end{figure}

As a first step we considered maps interacting with their nearest neighbors, corresponding to
$\mathcal N=4$ in (\ref{dyn1}, \ref{dyn2}). For various values of $g$ the characteristic length
$R$ was measured as a function of time: $R$ saturates for weak couplings at values small
compared to the lattice size. For larger couplings it shows scaling behaviour and one gets
complete phase separation in the spin variables. The value of $g$ that discriminates between
these two regimes is $g \simeq 0.03$; its precise evaluation is beyond the purposes of this
paper and will be presented elsewhere. Figures \ref{snap4f} and \ref{R4f} illustrate these
changes.
 Fitting late times growth in the phase separation region with the law $R(t)\sim At^z$
one gets the value $z=0.34 \pm 0.01$, in agreement with the result found by \cite{Koc-Cha} in
the case of 8 neighbors and with the Model B class of universality expectation
(Lifshitz-Slyozov law). For higher values of the coupling one could expect a faster transition
to the coarsening regime; however one finds a metastability region between $g=0.07$ and
$g=0.125$. Studying $R$ as a function of time one finds at small values of $R$ a plateau whose
extension grows with $g$. Between $g=0.120$ and $g=0.125$ the system gets trapped into blocked
configurations with interfaces between equal phase domains completely pinned. At $g=0.127$ the
behaviour of $R(t)$ suddenly changes, the plateau being reduced to a flex point; eventually,
for $g \gtrsim 0.128$ in few time - steps the lattice reaches a completely ordered
antiferromagnetic state (see fig.~\ref{n4g0128}(a)).
 Antiferromagnetic (AF) ordering at large
$g$ is a normal phenomenon: for $g>\frac{1}{\mathcal N}$ a map at time $t+1$ gets a
contribution with changed sign from its value at time $t$ and configurations with aligned spins
become unstable. This happens also in the case of CMLs with non conserved order parameter, i.
e. in absence of the second step of the dynamics. However, in the present case the AF ordering
is extremely precocious. The reason for this behaviour is the second redistributive step
(\ref{dyn2}) of the conservative dynamics. Let us focus our attention on site $i$ and let us
suppose, for example, that the first step (\ref{dyn1}) increases, in average, the values of its
neighbor maps. It is evident that, in the second step (\ref{dyn2}), this increment will result
in a negative contribution to the $x_i$ variable. This fact produces an effective
antiferromagnetic coupling. As we said, at $g=0.128$ the system quickly reaches
antiferromagnetic ordering, and this phenomenon is preceded by a blocked phase and by
metastability, which can be considered as pre-transitional effects. In fact, for lower values
of $g$, one can see that the domains of aligned spins have a checkerboard structure in which
values of maps belonging to various peaks of the asymptotic probability distribution functions
(PDFs) alternate (see fig.~\ref{n4g0127}).
 The transition to the AF ordering occurs because,
for $g \gtrsim 0.128$, the asymptotic PDF reduces to two narrow peaks of opposite sign (see
fig.~\ref{n4g0128}(b)). The AF ordering appears suddenly, the antiferromagnetic correlation
function displaying a jump after few tens of time steps. This ordering process survives for
higher values of $g$; subsequently the dynamics slows down until, by increasing $g$, the system
evolves towards blocked configurations. As many characteristics of this behaviour are similar
to the case of $\mathcal N=8$, we defer a more detailed analysis and discussion. Indeed, a new
interesting phenomenology comes out from the study of this same system when "laplacians" are
discretized using a higher number of neighbors, and it confirms the role of the second step of
the dynamics as generator of an effective AF coupling .

\section{Nearest and next to nearest neighbors simulations}
For small $g$ the behaviour of the CML in the case of eight neighbors is not substantially
different from the case of four neighbors, apart from the fact that the scale of $g$ is
smaller. For $g<0.007$ no phase separation is attained and the system gets blocked when equal
spin domains reach the dimension of few lattice spacings. For $g$ between $0.007$ and $0.110$
the growth of opposite phases domains does not stop and one measures a growth exponent of
$\frac{1}{3}$ at late times. However, for intermediate times scales, in the interior of equal
spin domains one can note the formation of a new sub arrangement of the maps, which corresponds
to stripe-ordering. The stripes in each domain correspond to peaks in PDFs belonging to the
same attractor of the map (\ref{map}).
\begin{figure}
\begin{center}
\includegraphics*[width=14cm]{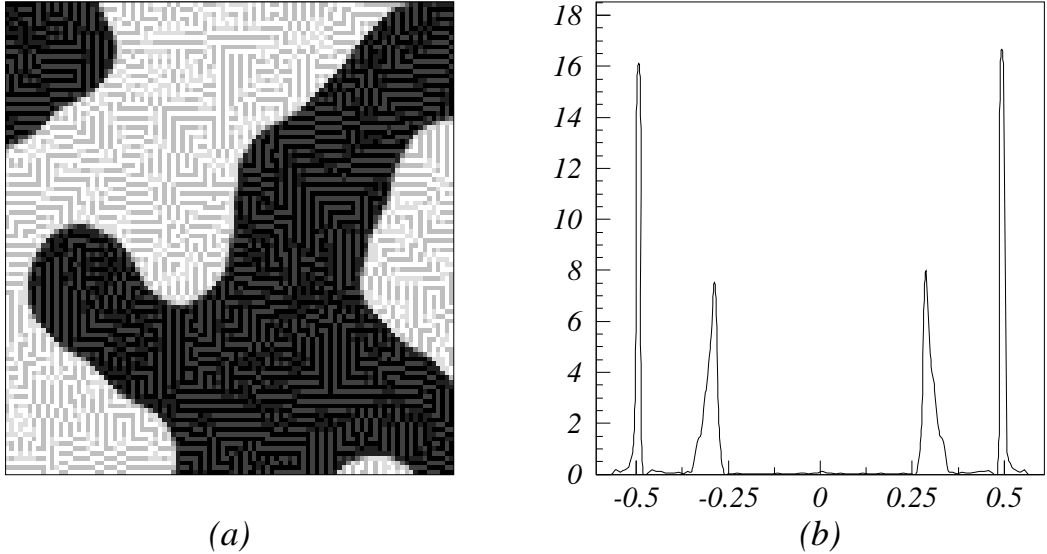}
\caption{The underlying striped structure for a CML with eight neighbors
$g=0.065$: a) a magnified portion ($100 \times 100$) of a $256 \times 256$
lattice after $10^5$ time steps, b) the probability distribution function for
the same lattice. The full range of maps' values has been colored with a 16
gray levels scale from black to white.} \label{n8g0065}
\end{center}
\end{figure}
 Fig.~\ref{n8g0065} shows for $g=0.065$ a snapshot of the
lattice and the sites PDF after $100000$ time steps. For $g \lesssim 0.110$ the competition
between ferromagnetic and lamellar ordering processes is resolved in favor of the former, but
this situation reverses for higher values of $g$. In this case the dynamics is essentially
composed by two stages: starting from a random configuration (with zero average) of the map
variables there is initially the formation of little striped domains oriented in the two
directions of the lattice, then the motion of the domains walls produces a coarsening process.

To study the growth of these striped domains we have considered, following
\cite{Sad-Bin,Cir-Gon-Str}, a new order parameter and two related correlation functions.
These articles study the growth processes of striped domains in the
superantiferromagnetic (SAF) phase of the Ising model with nearest and next to nearest
neighbors interactions. We recall that the SAF phase, in which the ground state is
four-fold degenerate and consists in alternate up and down spins rows or columns, is
related to the existence of an antiferromagnetic coupling between next to nearest
neighbors. In fact, if we call $J_1$ and $J_2$ the couplings between nearest neighbors
and between next to nearest neighbors respectively, the SAF phase corresponds to
$|J_1|<2|J_2|$ and $J_2>0$. The existence of striped domains is therefore another
corroboration for the existence of an effective antiferromagnetic coupling in the model
here presented. Let us divide the lattice in $2 \times 2$ cells and consider  a
two-components local order parameter
\begin{equation}\label{ord-par }
 \bf{\Psi}^{\vec\alpha}=\left(\begin{array}{c}
   \psi_1^{\vec\alpha} \\
   \psi_2^{\vec\alpha} \
 \end{array} \right)
\end{equation}
defined in each cell $\vec\alpha$ in the following way:
\begin{equation}\label{psi12}
 \psi_1^{\vec\alpha}=\sigma_1^{\vec\alpha}+\sigma_2^{\vec\alpha}-\sigma_3^{\vec\alpha}-\sigma_4^{\vec\alpha},
 \; \;  \psi_2^{\vec\alpha}=\sigma_1^{\vec\alpha}-\sigma_2^{\vec\alpha}-\sigma_3^{\vec\alpha}+\sigma_4^{\vec\alpha}
\end{equation}
where $\sigma_i^{\vec\alpha}$ are the clockwise ordered spins of the cell $\vec\alpha$. This
order parameter allows a univocal labelling of the four SAF phase ground states. $\bf{\Psi}$
allows the introduction of two correlation functions \cite{Cir-Gon-Str}:
\begin{eqnarray}\label{gamma_rl}
  \Gamma_\ell(r,t)&=&\frac{1}{2}\langle \psi_1^{\vec\alpha} \psi_1^{\vec\alpha+ r
  \vec x}\rangle+\frac{1}{2}\langle \psi_2^{\vec\alpha} \psi_2^{\vec\alpha+ r
  \vec y}\rangle,\\
  \Gamma_t(r,t)&=&\frac{1}{2}\langle \psi_1^{\vec\alpha} \psi_1^{\vec\alpha+ r
  \vec y}\rangle+\frac{1}{2}\langle \psi_2^{\vec\alpha} \psi_2^{\vec\alpha+ r
  \vec x}\rangle,
\end{eqnarray}
where $\vec x$ and $\vec y$ are  unit vectors in the $x$ and $y$ directions and
$\langle...\rangle$ indicates the average over lattice cells and different initial conditions.
$\Gamma_\ell$ and $\Gamma_t$ measure respectively the correlation properties in the direction
where the spins are aligned and in the direction where the spins are alternate.
\begin{figure}
\begin{center}
\includegraphics*[width=14cm]{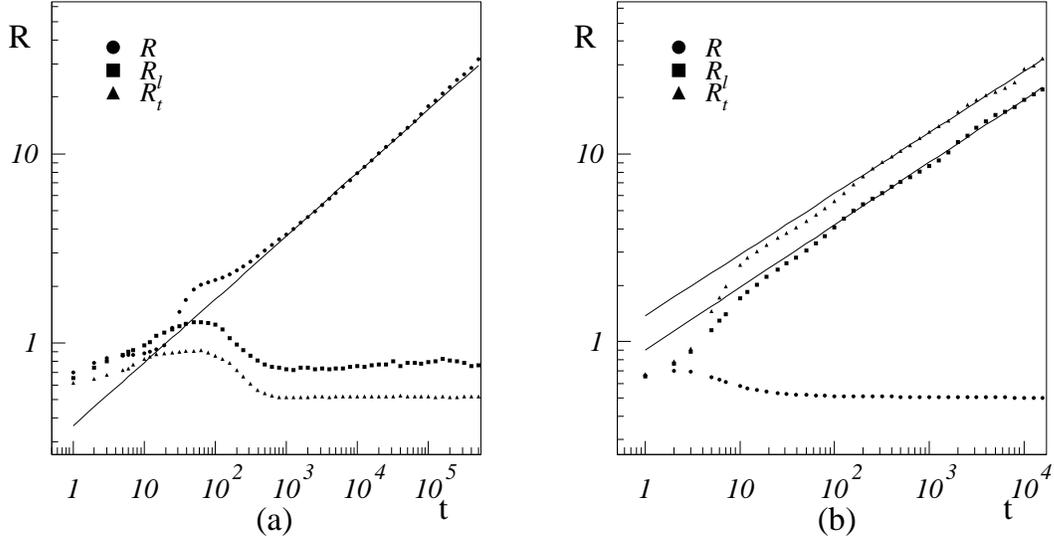}
\caption{a) The three measured lengths $R$ (circles), $R_l$ (squares), $R_t$
(triangles) as a function of time at $g=0.105$; the solid lines is a best fit
to the late time growth with power law $R(t)= At^z$ with $z=0.334$. b) The same
quantities at $g=0.111$; a power law fit gives in this case $z=0.337$ for $R_l$
and $z=0.326$ for $R_t$.} \label{n8rrr}
\end{center}
\end{figure}

\begin{figure}
\begin{center}
\includegraphics*[width=14cm]{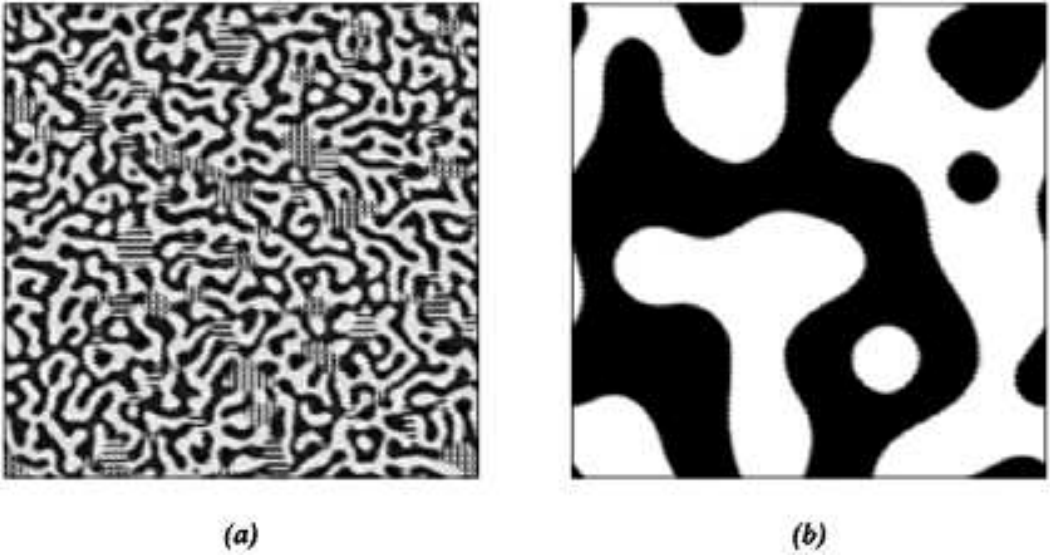}
\caption{Snapshots of a $256\times 256$ sites CML with eight neighbors for
$g=0.105$ (a) after $100$ and (b) after $50000$ time steps. The full range of
maps' values has been colored with a 16 gray levels scale from black to white.}
\label{n8g0105}
\end{center}
\end{figure}

\begin{figure}
\begin{center}
\includegraphics*[width=14cm]{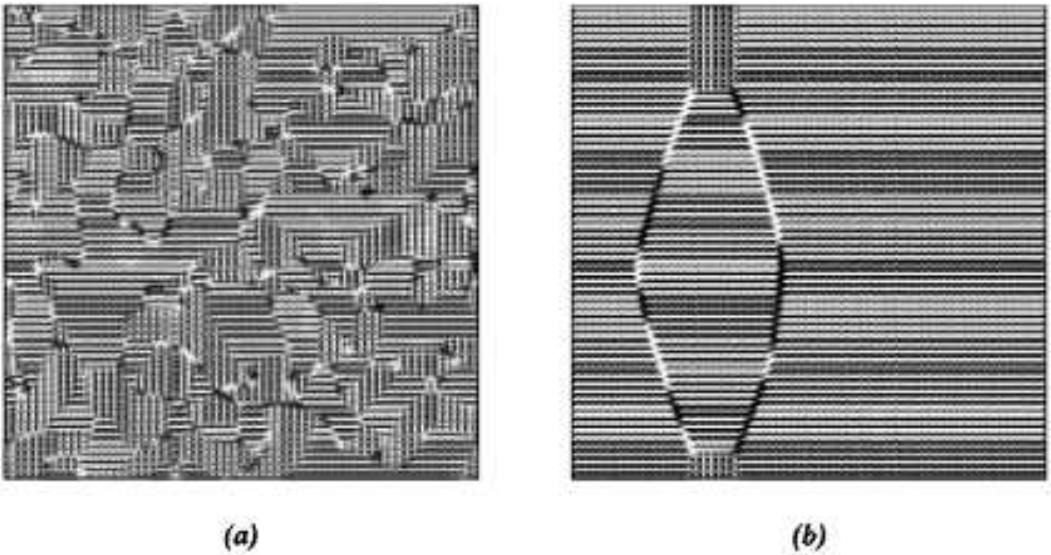}
\caption{Snapshots of a $256\times 256$ sites CML with eight neighbors for
$g=0.111$ (a) after $100$ and (b) after $50000$ time steps. The full range of
maps' values has been colored with a 16 gray levels scale from black to white.}
\label{n8g0111}
\end{center}
\end{figure}

Similarly to $C(r,t)$, it is possible to get from the measurement of $\Gamma_\ell(r,t)$ and
$\Gamma_t(r,t)$ two characteristic lengths $R_\ell(t)$ and $R_t(t)$ and to point out the
competition between the ordering processes. Figure \ref{n8rrr}(a) shows $R$, $R_\ell$ and $R_t$
as a function of time at $g=0.105$. There is an initial increase of all these three lengths,
then, after few dozens of time steps, the ferromagnetic ordering prevails, $R_\ell$ and $R_t$
go to zero and the growth of $R$ goes on, reaching the scaling regime power law with exponent
$\frac{1}{3}$. For higher values of $g$ the situation reverses. Fig.~\ref{n8rrr}(b) shows what
happens at $g=0.111$; $R_\ell$ and $R_t$ get quickly the Lifshitz-Slyozov law regime. These two
evolutions are also represented in fig.~\ref{n8g0105} and fig.~\ref{n8g0111} through two
snapshots of the lattice at $t=100$ and $t=50000$ for these two values of $g$.
 Another
interesting feature is the presence of an anisotropic growth of domains in the SAF phase
similar to that measured by \cite{Cir-Gon-Str}: $R_\ell$ is always greater than $R_t$, as long
as we are far enough from the formation of a single domain covering the whole lattice. This
happens, as reported in \cite{Cir-Gon-Str}, when the coupling between nearest neighbors is
ferromagnetic, while the reverse is true when this coupling changes sign.

\begin{figure}
\begin{center}
\includegraphics*[width=14cm]{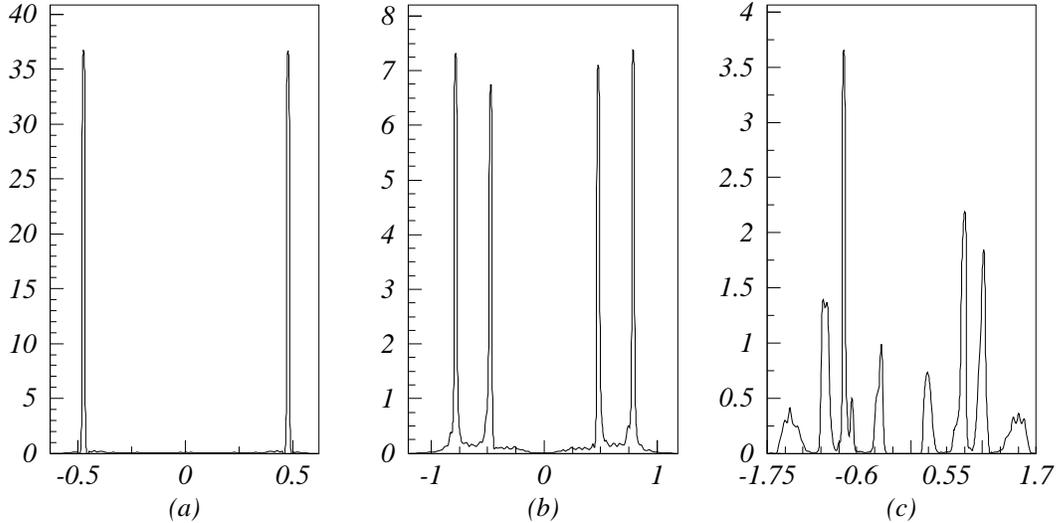}
\caption{Probability distribution functions for CMLs with eight neighbors after
$10^5$ time steps for $g=0.13$ (a), $g=0.20$ (b), $g=0.30$ (a).}
\label{n8peaks}
\end{center}
\end{figure}

This situation, in which one has complete striped ordering, persists increasing $g$ until $g
\simeq 0.13$. For higher values the incapacity of the system to get complete phase separation
couples to a behaviour of the maps' asymptotic PDFs that have characteristics similar to the
case ${\mathcal N} =4$. In fact, while at $g \simeq 0.13$, e. g., PDFs display two narrow peaks
at opposite values of the $x$ variable (fig.~\ref{n8peaks}(a)), increasing $g$ they display
first the broadening of these peaks, then the birth of two new peaks at symmetric $x$ positions
(fig.~\ref{n8peaks}(b)). A further increase of $g$ gives rise to an increasing number of peaks
(fig.~\ref{n8peaks}(c)).

At this point, order parameters like those borrowed from the Ising model, the ferromagnetic,
antiferromagnetic and lamellar ones, become inadequate to describe the dynamics of these
systems. Even when the lengths related to these order parameters have a monotonous behaviour,
one finds growth exponents different from the expected value of $\frac{1}{3}$. This exponent is
related to the mechanism decreasing the curvature between domains in which the order parameter
has an homogeneous value and to the conservation law. It is worthy to mention that the systems
we are studying satisfy, in this regime, the conservation law of a quantity that is no longer
the true order parameter for the coarsening process.

\section{Conclusions}

In this paper some results on the time evolution of lattices of coupled maps (to which an Ising
spin can be associated) in presence of a conservation law have been presented. Studying both
the cases of nearest and next to nearest neighbors, intervals of the coupling have been located
in which the coarsening process is similar to the classical growth phenomena described by model
B, confirming that also deterministic systems like CMLs are able to show similar processes of
phase separation. It has been pointed out an interesting feature, i.e. the effective
antiferromagnetic coupling rising from the conservation law constraint. Normally the study of
these systems has been limited to low values of the coupling $g$ and antiferromagnetic effects
have not been considered. The present analysis shows that they give rise to an interesting
phenomenology, including a scaling region and a phase diagram which resembles in some case the
Ising model. For example, for $\mathcal N =8$, starting from a ferromagnetic phase, one can go,
increasing the coupling value, to a Superantiferromagnetic phase through a paramagnetic one.
However, particularly for strong couplings, we have shown the greater complexity of these
models. This complexity cannot be described by concepts and quantities transferred from the
study of spin statistical models.

Furthermore this study confirms the importance of theoretical investigations of the asymptotic
probability distribution functions \cite{los-mac,cha-los} with respect to the ordering process.
With regard to this, one cannot exclude that, after a time longer than the one used in these
simulations, eight neighbors CMLs evolve definitely towards SAF configurations and their PDFs
towards two peaks structure also in the case of strong coupling.

Finally, in this analysis the parameter in (\ref{dyn2}), corresponding to the mobility
coefficient in the Cahn - Hilliard model, has the fixed value $1/\mathcal N$, so that, with the
subsequent sum, makes an average of the neighbors' value increments. Exploring the role of this
parameter could be the subject of future investigations on these systems.


\bibliographystyle{unsrt}
\bibliography{bibfile}

\end{document}